# EVIDENCE FOR HCO$^+$ INFALL TOWARD T TAURI?

Huib Jan van Langevelde [1]
Sterrewacht Leiden, P.O. Box 9513, 2300 RA Leiden, the Netherlands

Ewine F. van Dishoeck
Sterrewacht Leiden, P.O. Box 9513, 2300 RA Leiden, the Netherlands

Geoffrey A. Blake
Division of Geological and Planetary Sciences,
California Institute of Technology 170-25, Pasadena, CA 91125

**ABSTRACT.** High spectral and spatial resolution observations of the HCO$^+$ molecule toward the T Tau binary system obtained with the Owens Valley Millimeter Array reveal a broad emission line profile ($V_{\rm LSR}$ =7.2 km s$^{-1}$, $\Delta V$=2.0 km s$^{-1}$) upon which a narrow, redshifted absorption feature ($V_{\rm LSR}$ = 8.5 km s$^{-1}$, $\Delta V \approx$0.5 km s$^{-1}$) is superposed. One possible interpretation of the absorption is that it arises from infall of molecular cloud envelope material onto the immediate circumbinary region, which would imply that large scale dynamic accretion processes are still active well into the optically visible stages of star formation. Provided the mass is deposited at radii sufficiently close to the young stars, estimates of the accretion rates indicate that infall may add enough material to the system to account for the FU Orionis type flares recently observed in T Tau in $\lesssim$ 100 yr.

*Subject headings:* Stars: T Tauri — Stars: pre–main–sequence — Stars: Formation — ISM: molecules

[1] Currently at Joint Institute for VLBI in Europe, Radiosterrenwacht Dwingeloo, Postbus 2, 7990 AA Dwingeloo, the Netherlands

1. INTRODUCTION

T Tauri stars are the first optically visible stage of low–mass star formation. In the standard scenario of Shu et al. (1987) a T Tauri star emerges from its natal cocoon of gas and dust when bipolar outflow or other dissipative processes have had enough time to clear the surroundings. At this stage, which occurs about $10^6$ yr after the onset of proto–stellar collapse, the star is still surrounded by dense circumstellar material, possibly in the form of a centrifugally supported accretion disk (for recent observational reviews, see Basri & Bertout 1993; Sargent & Welch 1993). Since such disks are thought to be the predecessors of planetary systems, detailed information on their evolution is warranted. Of particular interest is the question whether the accretion of matter from large distances onto the disk or young star is still occurring at this optically visible stage of the evolution (Galli & Shu 1993). Continued accumulation of material onto the disk from a surrounding envelope might lead to gravitational instabilities and FU Orionis–type outbursts, thereby influencing the early stages of dust grain coagulation and growth or other "planetary" formation processes (Hartmann et al. 1993). Although there is some recent observational evidence for collapse during the deeply embedded phase of low–mass star formation at $\lesssim 10^5$ yr (Zhou et al. 1993), as well as for collapse of large high–mass star forming regions as a whole (c.f. Carral & Welch 1992), little information is yet available on the later T Tauri phase. We present here kinematic evidence which could be interpreted as the infall of molecular cloud material well into the classical T Tauri stages of stellar formation.

T Tau itself, as the prototype of its class, has been extremely well studied from radio to ultraviolet wavelengths. The relatively close distance of the Taurus cloud ($\approx$ 140 pc, Elias 1978) facilitates the separation of the various stellar and circumstellar components. As is quite common for young stars, T Tau is part of a binary system, with a separation of only 100 AU ($0\rlap.{''}73$) in the N–S direction (Ghez et al. 1991). The primary star, called T Tau N, is optically visible and shows all the stellar characteristics of its type. The companion, called T Tau S, is only observable at infrared (IR) or longer wavelengths, suggesting that it is still surrounded by an envelope of gas. Its spectral energy distribution, however, indicates that its age cannot be much less than that of the primary.

The existence of circumstellar material around T Tau has been deduced from IR and submillimeter observations of the continuum radiation from dust (Weintraub et al. 1989$b$, 1992; Beckwith et al. 1990; Beckwith & Sargent 1991), and material on Keplerian orbits encompassing both young stellar objects has been suggested from millimeter–wave interferometric observations of CO (Weintraub et al. 1989$a$). The evidence for material inside the orbit of the companion is ambiguous since the system orientation (nearly pole on, Herbst et al. 1986) is such that the bipolar outflow along the line of sight confuses the kinematic picture. On the other hand, this geometry has the advantage that the full potential disk/envelope area is seen, resulting in larger fluxes.

We report here observations of the $HCO^+$ ion, which provide insight into the questions of infall, disk/envelope truncation, and outflow. In particular, $HCO^+$ is a better tracer than CO of any infall of material because of its much larger dipole moment, as has been demonstrated for several star–forming regions in which molecular lines — quite often $HCO^+$ $J=1\rightarrow 0$ — appear self–absorbed (Welch et al. 1987; Zhou et al. 1993; Carral & Welch 1992; Rudolph et al. 1990).

## 2. OBSERVATIONS AND LINE SHAPES

Fig. 1a,b presents Owens Valley Radio Observatory (OVRO) Millimeter Array HCO$^+$ J=1→0 spectra in correlator backend setups of 125 kHz (or 0.42 km s$^{-1}$) and 31.25 kHz (0.105 km s$^{-1}$) resolution from a 7″ × 7″ box (≈ the naturally weighted beam size) centered on the position of T Tau. Data were obtained in five array configurations between 25 April 1992 and 3 July 1993. Calibration was performed with 0420−014 serving as the phase calibrator and 3C273 as the amplitude calibrator. Similar observations were acquired for the $^{13}$CO J=1→0 and C$^{18}$O J=1→0 transitions between January and June 1993. The spectral resolutions are 125 kHz for both transitions, and the synthesized beam is ≈ 5″.8 × 4″.9.

The OVRO HCO$^+$ spectrum shows a broad emission feature (FWHM $\Delta V$ ≈ 2.0 km s$^{-1}$) with a maximum at $V_{\rm LSR}$ = 7.2 km s$^{-1}$. The low resolution spectrum is quite asymmetric with a sharp edge at the redshifted side, reaching a minimum at ≈ 8.7 km s$^{-1}$. Such a blue/redshifted asymmetry was noticed earlier by Weintraub et al. (1989a) in their $^{12}$CO J=1→0 OVRO spectra. The high-resolution spectrum demonstrates that the asymmetry is caused by a narrow ($\Delta V$ ≈ 0.5 km s$^{-1}$) dip at $V_{\rm LSR}$ ≈ 8.5 km s$^{-1}$. Fig. 1c,d also includes T Tau and (0″, -30″) offset HCO$^+$ J=1→0 spectra observed with the 30m IRAM telescope in May 1991. The IRAM on-source spectrum in a 25″ beam is considerably different from that obtained at OVRO; it is dominated by a strong narrow ($\Delta V$ ≈ 1.2 km s$^{-1}$) component at $V_{\rm LSR}$ ≈ 8.3 km s$^{-1}$ that is resolved out by the interferometer, so that it must be extended over more than 30″. This is consistent with the observed narrow emission at the offset position, for which $V_{\rm LSR}$ = 8.5 km s$^{-1}$. Maps in HCO$^+$ J=1→0 and $^{12}$CO J=3→2 around T Tau show strong narrow emission over a region of about 40″, with weaker emission present over a scale of several arcmin. We identify this component with an extended molecular envelope around the T Tau system on scales of order ≈ 6000 AU, with additional emission on larger scales.

The second component in the IRAM T Tau J=1→0 spectrum is well fit by a Gaussian profile with $V_{\rm LSR}$ ≈ 7.5 km s$^{-1}$, $\Delta V$ ≈ 2.8 km s$^{-1}$. This emission resembles that seen by OVRO, and must be due to a relatively compact object. It becomes more prominent in the higher excitation J=3→2 and J=4→3 lines observed with the James Clerk Maxwell Telescope [1] (JCMT) in August 1993 (Fig. 1e,f) in a 19″ and 14″ beam, respectively. The J=3→2 line is dominated by emission at $V_{\rm LSR}$ ≈ 7.8 km s$^{-1}$, but still shows considerable narrow emission at $V_{\rm LSR}$ ≈ 8.3 km s$^{-1}$. The J=4→3 line, however, is nearly symmetric with $\Delta V$ ≈ 3.2 km s$^{-1}$ at $V_{\rm LSR}$ ≈ 7.6 km s$^{-1}$. Clearly the compact object at $V_{\rm LSR}$ ≈ 7.5 km s$^{-1}$ is warm and dense given the conditions required to excite HCO$^+$ to these levels ($n_{\rm H_2} > 10^6$ cm$^{-3}$, $T_{\rm kin} > 30$ K). Limited mapping in the HCO$^+$ J=3→2 and J=4→3 lines shows that the narrow emission at $V_{\rm LSR}$ ≈ 8.3 km s$^{-1}$ is extended over about 20″ − 30″ with densities of order $10^5$ cm$^{-3}$. At a 120″ offset position, the density has decreased to about $2 \times 10^4$ cm$^{-3}$, typical of the general Taurus cloud.

Inspection of the OVRO HCO$^+$ J=1→0 map, presented in Fig. 2, shows that most of the emission is contained in a source with a deconvolved size of 9″.2 × 7″.1 (1300 × 1000 AU). The maximum of the integrated emission is slightly offset (by 2″.5) to the NW compared with the broad band continuum emission. This may be related to the

---

[1] The JCMT is operated by the Royal Observatory, Edinburgh, on behalf of the Science and Engineering Research Council of the UK, the Netherlands Organisation for Scientific Research and the National Research Council of Canada.

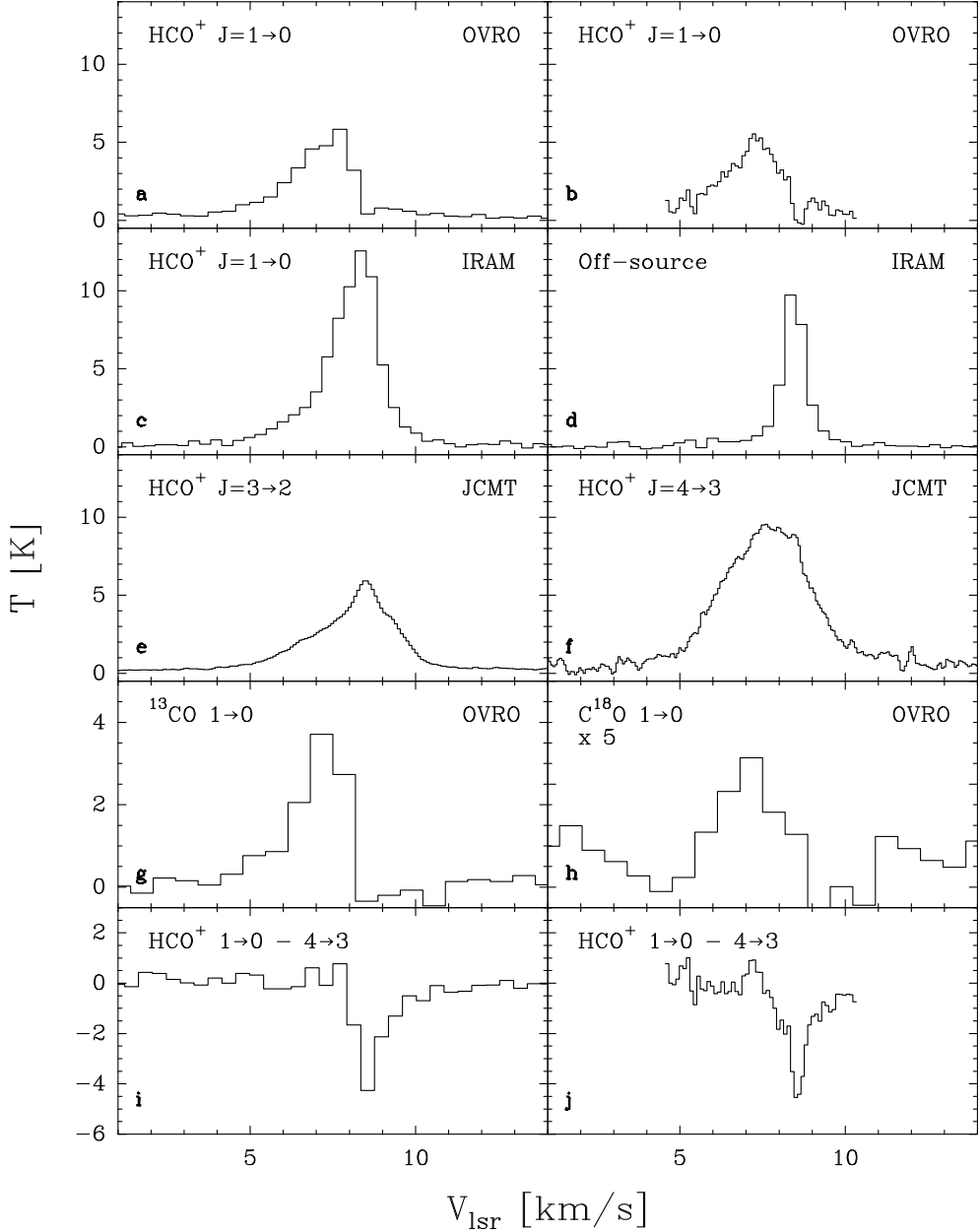

**1.** Spectra of T Tau. **a.** $HCO^+$ $J=1\to 0$ low resolution spectrum obtained with OVRO in a $7''\times 7''$ box around the measured central position ($\alpha,\delta$ (1950) = $4^h19^m4^s.12$, $19°25'5''.7$). **b.** Same as a. but for the high resolution backend. **c.** $HCO^+$ $J=1\to 0$ line from the IRAM 30m ($25''$ beam). **d.** $HCO^+$ $J=1\to 0$ line from the IRAM 30m at an off-source position ($0,-30''$). **e.** $HCO^+$ $J=3\to 2$ observed with the JCMT ($19''$ beam). **f.** $HCO^+$ $J=4\to 3$ observed with the JCMT ($14''$ beam). **g.** OVRO $^{13}CO$ $J=1\to 0$ spectrum. **h.** $C^{18}O$ $J=1\to 0$ transition from OVRO, scaled by a factor 5 in comparison to $^{13}CO$. **i,j.** OVRO low and high resolution $HCO^+$ $J=1\to 0$ spectra with the $HCO^+$ $J=4\to 3$ JCMT profile subtracted, scaled to match the blue wing of the OVRO data.

asymmetry in the line profile; strong blueshifted emission occurs in the NW, whereas the weak redshifted emission is to the E.

Although the interferometer $HCO^+$ $J=1\to 0$ spectrum does not show narrow emission at $V_{LSR} \approx 8.3$ km s$^{-1}$, it does have a narrow dip —possibly absorption— at $V_{LSR} \approx 8.5$ km s$^{-1}$. This feature is not seen in the single dish $J=1\to 0$ spectrum because it is

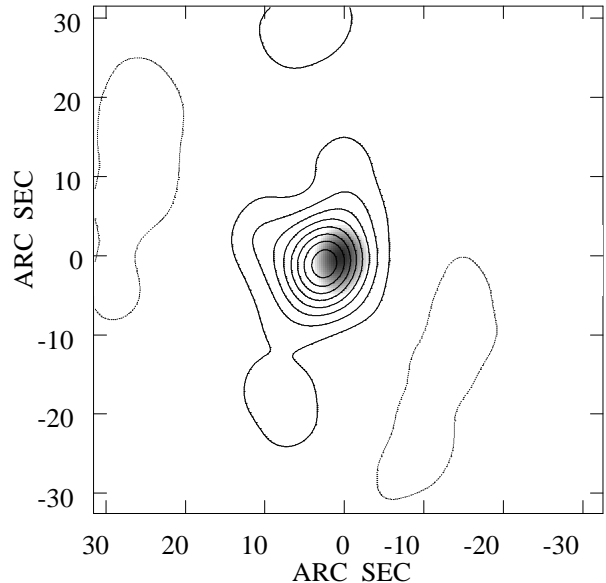

2. Map of the integrated HCO$^+$ profile (contours) superimposed on a grey scale image of the continuum. The contours are drawn at 2.0 K km s$^{-1}$ intervals starting at 2.0 K km s$^{-1}$.

overwhelmed by the strong and extended narrow emission. A similar dip also appears present in the OVRO $^{13}$CO $J=1\rightarrow 0$ spectrum at $V_{\rm LSR} \approx 8.5$ km s$^{-1}$, but not in that of C$^{18}$O $J=1\rightarrow 0$ (see Fig. 1g,h), although the C$^{18}$O is quite weak. These lines show emission with $\Delta V \approx 1.5$ and 1.7 km s$^{-1}$, respectively, comparable to the HCO$^+$ $J=1\rightarrow 0$ interferometric profile.

A better view of the absorption can be obtained by assuming that the intrinsic profile of the $J=4\rightarrow 3$ line is the same as that of the $J=1\rightarrow 0$ line seen in the interferometer, scaling it to match the blue wing and subtracting it. The result is presented in Fig. 1i,j. Note that the absorption can only be observed at the velocities where background emission is available with good signal to noise ratios, implying that velocities to the red of $V_{\rm LSR} \approx 11$ km s$^{-1}$ are poorly sampled. Careful analysis of the maps of the HCO$^+$ and $^{13}$CO emission at 8.5 km s$^{-1}$ rule out an instrumental origin for the dip in the spectrum.

## 3. INFALL TOWARD T TAU?

What physical phenomena could give rise to these line shapes? Possibilities include rotation in a circumbinary disk or envelope, outflow, or infall. In the following, we will examine each of these options and conclude that the broad emission is due to circumbinary or accreting matter close to the stars, and that infall from larger radii is a plausible explanation of the dip in the OVRO HCO$^+$ and $^{13}$CO spectra due to absorption.

One possible alternative explanation for the dip in the spectrum would be an asymmetric density distribution or a hole in a circumbinary disk caused, for example, by the orbiting companion. Similarly, two very compact circumstellar disks, one around the primary at $V_{\rm LSR} \approx 7.5$ km s$^{-1}$ and one around the companion at $V_{\rm LSR} \approx 9.2$ km s$^{-1}$, could perhaps mimic the HCO$^+$ $J=1\rightarrow 0$ profile. Such gaps and compact binary disks have been postulated on the basis of near– and mid–IR observations. However, the derived sizes of 10 AU and 50 AU (Koresko 1993) are much too small to be consistent with the observed flux, and because the feature is so narrow any asymmetry would have to occur at a very specific location. Finally, these models are also unable to explain the profiles of the higher excitation HCO$^+$ lines or that of C$^{18}$O, which show no dips.

Another possibility involves bipolar outflow, which is known to be present in this system (Edwards & Snell 1982; Levreault 1988a). Outflow is the likely cause of the low–intensity wings seen in the $J$=1→0 and $J$=3→2 spectra at > 2 km s$^{-1}$ from line center. Although a detailed model for this outflow is not available, it seems difficult to produce spectra with a high degree of symmetry in the higher excitation lines but asymmetric in the $J$=1→0 line. Further, only *blueshifted* absorption would be a natural phenomenon in an expanding outflow.

The most likely explanation of the dip in the spectrum is thus absorption by a foreground layer. Moreover, this model naturally explains the profiles of the other observed lines. We assume that the absorption occurs primarily against the circumbinary broad emission profile centered at $V_{\rm LSR} \approx 7.5$ km s$^{-1}$. This compact emission must clearly be associated with T Tau itself and arises from warm, dense gas on a 1000 AU scale. It could be either due to a large circumbinary disk similar to that of Weintraub et al. (1989a), or to the inner, dense part of an accretion envelope (Terebey et al. 1984; Kenyon et al. 1993). Detailed comparison of the observed emission profile with theoretical calculations will be presented in a later paper, but we stress that an unbiased evaluation of rotation versus accretion depends critically on accurate determinations of the system geometry.

Clearly, the absorbing material must be in front of T Tau, and because it is redshifted with respect to the central velocity, it must either be moving towards the disk, i.e., falling in, or the T Tau system has some net motion with respect to its parental cloud. Given the differing nature of these interpretations, it is important to localize the absorbing layer and to decipher whether it is gravitationally bound to T Tau itself. The absorbing material is apparently too cold and diffuse to provide considerable opacity in the higher HCO$^+$ $J$=3→2 and $J$=4→3 transitions, although emission from the extended background cloud could be filling in part of the absorption, in particular for the $J$=3→2 spectrum. From upper limits on the optical depths in these lines $T_{\rm kin} < 30$ K and $n_{\rm H_2} \lesssim 10^5$ cm$^{-3}$ is estimated.

The optical depth of the HCO$^+$ $J$=1→0 absorption is $\tau \approx 3$, which implies $N_{\rm HCO^+} \approx (2-4) \times 10^{12}$ cm$^{-2}$ for the conditions derived above. The typical abundance of HCO$^+$ in the Taurus cloud is $(6-8) \times 10^{-9}$ (Guélin et al. 1982; Ohishi et al. 1992), resulting in a total H$_2$ column density of $(3-6) \times 10^{20}$ cm$^{-2}$. This column corresponds to a size of $\approx 1000 - 2000$ AU, which is comfortably larger than the size of the warm and dense circumbinary material. It is likely to be a lower limit, since the HCO$^+$ abundance in such low column density gas may be substantially lower. Indeed, for an optical depth in $^{13}$CO $J$=1→0 of order 0.5 – 1, a $^{13}$CO column of $(2-4) \times 10^{15}$ cm$^{-2}$ is derived, corresponding to $N_{\rm H_2} \approx (1-2) \times 10^{21}$ cm$^{-2}$, or $A_V \approx 1-3$ mag. For spherical symmetry the derived accretion rate is $\approx 2 \times 10^{-7} M_\odot$ yr$^{-1}$, which is comparable to current estimates of the *mass loss* rate, $1.3 \times 10^{-7} M_\odot$ yr$^{-1}$ (Levreault 1988b). This number is a lower limit to the actual infall rate. Any warmer, denser gas entrained in infall closer to the binary would not contribute substantially to the $J$=1→0 absorption, and the HCO$^+$ abundance might be lower at these higher densities than in the Taurus molecular cloud. The redshifted absorption "tail" near $V_{\rm LSR} \sim 10-11$ km s$^{-1}$ (Figs. 1i,j) may arise in part from such gas.

The absorption interpretation can also successfully explain the results from the OVRO maps. If the main HCO$^+$ emission is from a circumbinary disk or accretion envelope, we expect symmetric emission around the stellar position, with one side having predominantly approaching and the other receding velocities. When considerable absorption occurs at red–shifted velocities, an asymmetric image is to be expected, with predominant blue–shifted velocities. This matches the observed spatial distribution (e.g., Fig. 2). The shapes of the $^{13}$CO and C$^{18}$O $J$=1→0 transitions can also be readily understood by recalling that CO has a very small dipole moment. As a result, the Einstein A coefficients of the CO isotopomers are factors of 1000 smaller than that of HCO$^+$. For $^{13}$CO, this reduction is partly offset by the fact that $N_{^{13}CO} \approx 200 - 1000 \times N_{HCO^+}$, so that some $^{13}$CO absorption occurs. However, C$^{18}$O is only a factor 20 – 100 more abundant, so that its optical depth should be very small, resulting in negligible absorption.

The size of the infalling envelope might be similar to that of the extended envelope of $\sim$ 6000 AU seen in the narrow emission at offset positions (Fig. 1c, see §2). However, for this extended emission, which was also detected in the higher excitation lines, a larger column density, temperature and density are required ($N_{HCO^+} \approx 2 \times 10^{13}$ cm$^{-2}$, $T_{kin} \approx 30$ K, $n_{H_2} \gtrsim 10^5$ cm$^{-3}$). With the above estimates of the HCO$^+$ abundance this would lead to $A_V \gtrsim 4$ mag. At the lower density 120″ offset position, the HCO$^+$ column density implies $A_V \approx 6$ mag. It seems plausible that at least half of this gas is on the far side of T Tau. The values of $A_V$ toward the stellar components are estimated to be 1.4 mag to T Tau N (Cohen & Kuhi 1979), and $A_V \approx 4.5$ mag to T Tau S (Ghez et al. 1991).

The HCO$^+$ seen in absorption could be in the colder, less dense outer region of the envelope around T Tau, since cold HCO$^+$ is a very efficient absorber. Emission from HCO$^+$ with the parameters of the absorbing component would be difficult to detect: a possible blueshifted counterpart falling in at the opposite side of T Tau would contribute only 0.5 K to the emission profile if $T_{kin} \sim$10 K and $n_{H_2} \sim 10^4$ cm$^{-3}$. Although there is no way to determine exactly where the absorbing material is located on this relatively short line of sight (140 pc), it is plausible that it is associated with the immediate surroundings of T Tau on the basis of the above arguments. The observation of self-absorbed higher excitation HCO$^+$ and CO lines at interferometric spatial resolution would further support the proposed infall model.

The estimated size of the absorbing layer is also comparable to that of the nebula found around T Tau from near–IR images of scattered light by Weintraub et al. (1992). This nebula provides a plausible origin of the infalling gas, and we note that both the size of the near–IR envelope and the HCO$^+$ distributions presented here are consistent with an infall scenario for the estimated system mass of 3 M$_\odot$ (Bertout 1983). Whether the material actually falls onto a compact disk around T Tau N, T Tau S or both cannot be distinguished from our observations, but it is likely that at least some of the material is falling onto the disk surrounding the infrared companion. This continued supply of material could cause the disk to become unstable after some period of time, which might in turn give rise to FU Orionis type outbursts. In fact, an infrared flare has recently been observed for T Tau S by Ghez et al. (1991). The mass loss of $\approx 3.6 \times 10^{-6}$ $M_\odot$ yr$^{-1}$ over the period of the flare of $\lesssim 1$ yr could have been accumulated in $\lesssim$ 100 years. The current accretion rate, as well as the observed velocity, indicate that this phase could be sustained for an additional several times $10^5$ years.

The authors would like to thank the JCMT, IRAM and OVRO staff. OVRO is operated under funding from the NSF, contract 90–16404. Frank Helmich is thanked for carrying out the JCMT observations. EvD and HJvL acknowledge support from the Netherlands Organization for Scientific Research (NWO), and GAB from the David and Lucille Packard and Alfred P. Sloan Foundations, as well as NASA grant NAGW-2297.

## 4. REFERENCES

Basri G. & Bertout C., in *Protostars & Planets III*, ed. E.H. Levy & J.I. Lunine (Univ. Arizona Press), p. 543.

Beckwith S.V.W., Sargent A.I., 1991, ApJ, 381, 250

Beckwith S.V.W., Sargent A.I., Chini R., Güsten R., 1990, AJ, 99, 924

Bertout C., 1983, A&A, 126, L1

Carral P., Welch W.J., 1992, ApJ, 385, 244

Cohen M., Kuhi L.V., 1979, ApJS, 41, 743

Edwards S., Snell R.L., 1982, ApJ, 151, 160

Elias J.H., 1978, ApJ, 224, 857

Galli D., Shu, F.H., 1993, ApJ, 417, 243

Ghez A.M., Neugebauer G., Gorham P.W., Haniff C.A., Kulkarni S.R., 1991, AJ, 102, 2066

Guélin M., Langer W.D., Wilson R.W., 1982, A&A, 107, 107

Hartmann L., Kenyon S., Hartigan P., in *Protostars & Planets III*, ed. E.H. Levy & J.I. Lunine (Univ. Arizona Press), p. 497.

Herbst W., Booth J.F., Chugainov P.F., Zajtseva G.V., Barksdale W., Covino E., Terranegra L., Vittone A., Vrba F., 1986, ApJ, 310, L71

Kenyon S.J., Calvet N., Hartmann L, 1993, ApJ, 414, 676

Koresko CD., 1993, Ph. D. Thesis, Cornell University.

Levreault R.M., 1988*a*, ApJS, 67, 283

Levreault R.M., 1988*b*, ApJ, 330, 897

Ohishi M., Irvine W.M., Kaifu N., 1992, in *Astrochemistry of cosmic Phenomena*, IAU 150, ed. P.D.Singh (Kluwer Dordrecht) p. 171

Rudolph A., Welch W.J., Palmer P., Dubrulle B., 1990, ApJ, 363, 528

Sargent, A.I. & Welch, W.J., 1993, ARA&A, 31, 297

Schwartz P.R., Simon T., Campbell R., 1986, ApJ, 303, 233

Shu, F.H., Adams, F.C., Lizano, S., 1987, ARA&A, 25, 23

Terebey S., Shu F.H., Cassen P., 1984, ApJ, 286, 529

Weintraub D.A., Masson C.R., Zuckerman B., 1989*a*, ApJ, 344, 915

Weintraub D.A., Sandell G., Duncan W.D., 1989*b*, ApJ, 340, 69

Weintraub D.A., Kastner J.H., Zuckerman B., Gatley I., 1992, ApJ, 391, 784

Welch W.J., Dreher J.W., Jackson J.M., Terebey S., Vogel S.N., 1987, Sci 238, 1550

Zhou, S.D, Evans N J, Kömpe C., Walmsley C.M., 1993, ApJ, 404, 232